\documentclass[12pt,twoside]{article}
\usepackage{amsmath}
\usepackage{amssymb}
\usepackage{array}
\usepackage{graphicx}
\usepackage{geometry}
\usepackage{graphics}
\usepackage{enumerate}
\usepackage[utf8]{inputenc}

\begin{document}

\begin{center}

{\bf Does SuperKamiokande Observe L\'evy  Flights of Solar Neutrinos?}

\vskip1cm

Arak M. Mathai\\
Department of Mathematics and Statistics\\
McGill University, Montreal, Canada\\
a.mathai@mcgill.ca\\

and\\

Hans J. Haubold\\
Office for Outer Space Affairs, United Nations,\\
Vienna International Centre, Vienna, Austria\\
hans.haubold@gmail.com\\
https://universeexplorer.org/

\end{center}

\vskip.5cm\noindent{\bf Abstract.}
The paper is utilyzing data from the SuperKamiokande solar neutrino detection experiment and analyses them by diffusion entropy analysis and standard deviation analysis to evaluate the scaling exponent of the probability density function. The result of analysis indicates that solar neutrinos are subject to L\'evy flights. The paper derives the probability density function and the governing fractional diffusion equation for solar neutrino L\'evy flights in terms of Fox's H-function. The conclusion of the paper is the question: Does SuperKamiokande Observe L\'evy  Flights of Solar Neutrinos?

\vskip.3cm\noindent{\bf Keywords:} solar neutrinos, SuperKamiokande experiment data, diffusion entropy analysis, standard deviation analysis, scaling exponent, neutrino probability density function, Fox H-function, fractional diffusion equation, L\'evy flights.

\vskip.3cm\noindent{\bf 1.\hskip.3cm Solar Neutrinos: SuperKamiokande Data}

\vskip.3cm\noindent Over the past 50 years, radio-chemical and real-time solar neutrino experiments have proven to be sensitive tools to test both astrophysical and elementary particle physics models and principles (Sakurai, 2018; Orebi Gann et al., 2021). Solar neutrino detectors (radio-chemical: Homestake, GALLEX + GNO, SAGE, real-time: SuperKamiokande, SNO, Borexino)  have demonstrated that the Sun is powered by thermonuclear fusion reactions. Today fluxes, particularly from the pp-chain have been measured: $pp$. $^7Be$, $pep$, $^8B$, and, $hep$. Experiments with solar neutrinos and reactor anti-neutrinos (KamLAND) have confirmed that solar neutrinos undergo flavor oscillations (Mikheyev-Smirnov-Wolfenstein (MSW) model). Results from solar neutrino experiments are consistent with the Mikheyev-Smirnov-Wolfenstein Large Mixing Angle (MSW-LMA) model, which predicts a transition from vacuum-dominated to matter-enhanced oscillations, resulting in an energy dependent electron neutrino survival probability.

\begin{figure}
\resizebox{12cm}{!}{\includegraphics{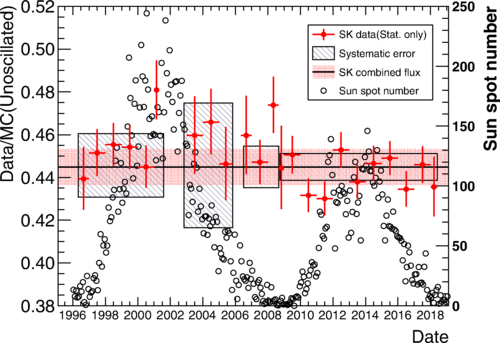}}
\caption{Yearly solar neutrino flux measured by SuperKamiokande. The redfilled circle points show the SuperKamiokande data with statistical uncertainty and the gray striped area show the systematic uncertainty for each phase. The horizontal black solid line (red shaded area) shows the combined value of measured flux (its combined uncertainty). The black-blank circle points show the sunspot numbers from 1996 to 2018 (Abe et al., 2024)}
\end{figure}

\begin{figure}
\resizebox{12cm}{!}{\includegraphics{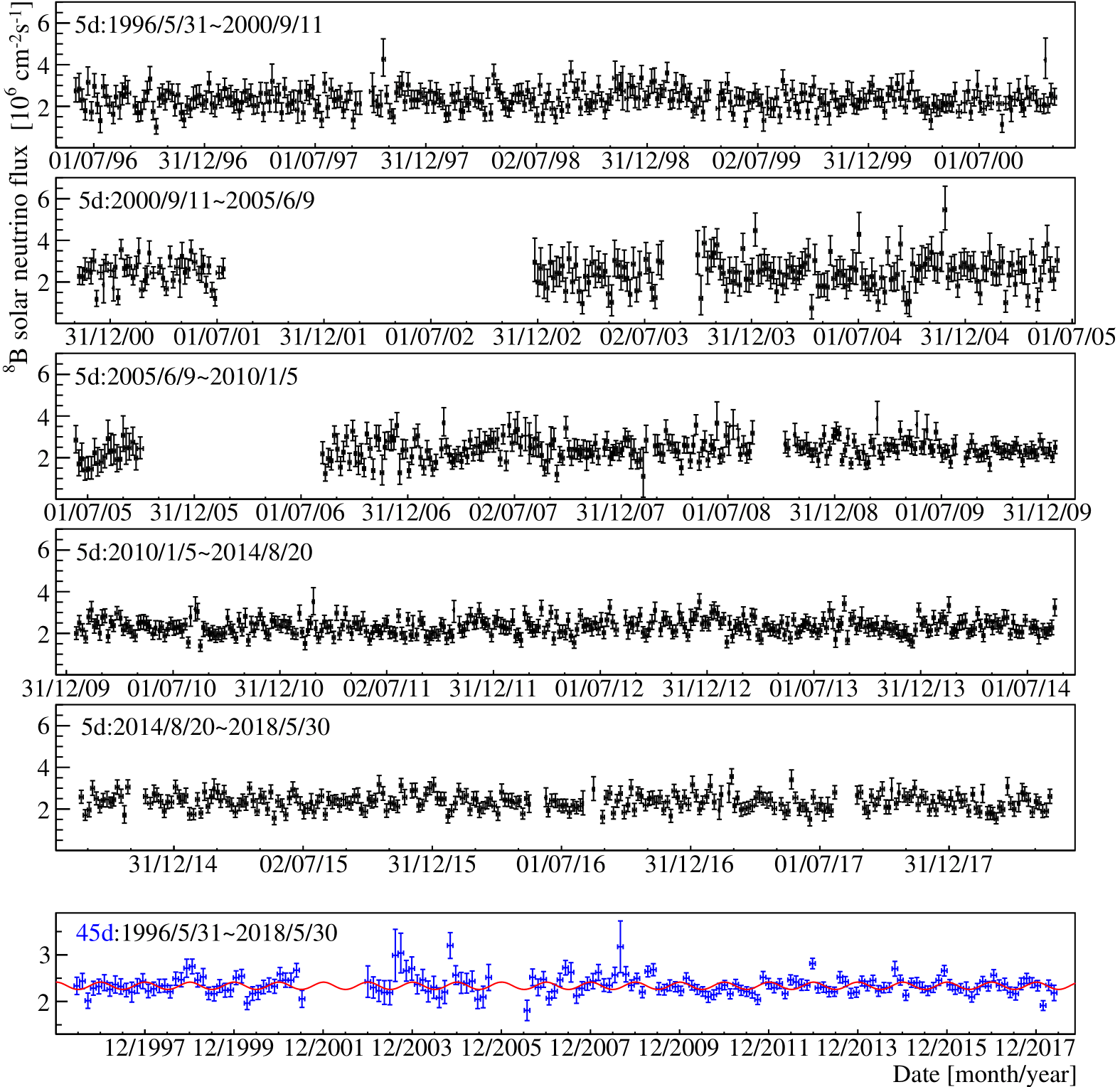}}
\caption{Measured $^8B$ solar neutrino fuxes for 5-day (top five panels, black data points) and 45-day (bottom panel, blue data points) intervals without 1/$R^2$ correction. The errors in the 5-day (the 45-day) plot are asymmetric (symmetric) errors of the average fluxes. The solid-red curve in the 45-day plot is the expected sinusoidal solar neutrino flux based on the elliptical orbit of the Earth Abe et al., 2023.}
\end{figure}

\vskip.3cm\noindent{\bf 2.\hskip.3cm Diffusion Entropy and Standard Deviation: Analysis}
\vskip.3cm\noindent
For all radio-chemical and real-time solar neutrino experiments, periodic variation in the detected solar neutrino fluxes have been reported, based mainly on Fourier and wavelet analysis methods (standard deviation analysis). Other attempts to analyze the same data sets, particularly undertaken by the experimental collaborations of real-time solar neutrino experiments themselves, have failed to find evidence for such variations of the solar neutrino flux over time (Abe et al., 2023). Periodicities in the solar neutrino fluxes, if confirmed, could provide evidence for new solar, nuclear, or neutrino physics beyond the commonly accepted physics of vacuum-dominated and matter-enhanced oscillations of massive neutrinos (MSW model) that is, after 50 years of solar neutrino experiment and theory, considered to be the ultimate solution to the solar neutrino problem (Figure 1).\\
Specifically, subsequent to the analysis made by the SuperKamiokande collaboration, the SNO experiment collaboration has painstakingly searched for evidence of time variability at periods ranging from 10 $years$ down to 10 $minutes$. SNO has found no indications for any time variability of the $^8B$ flux at any timescale, including in the frequency window in which $g$-mode oscillations of the solar core might be expected to occur. Despite large efforts to utilize helio-seismology and helio-neutrinospectroscopy, at present time there is no conclusive evidence in terms of physics for time variability of the solar neutrino fluxes from any solar neutrino experiment. If such a variability over time would be discovered, a mechanism for a chronometer for solar variability could be proposed based on relations between properties of thermonuclear fusion and g-modes (Buldgen et al., 2024; Sturrock et al., 2021).\\
All above findings encouraged the conclusion that Fourier and wavelet analysis, which are based upon the analysis of the variance of the respective time series (standard deviation analysis: SDA) should be complemented by the utilization of diffusion entropy analysis (DEA), which measures the scaling of the probability density function (pdf) of the diffusion process generated by the time series thought of as the physical source of fluctuations (Scafetta, 2010). For this analysis, we have used the publicly available data of SuperKamiokande-I (1996-05-31 - 2001-07-15) and SuperKamiokande-II (2002-12-10 - 2005-10-06) (see Figure 2) (Yoo et al., 2003: Cravens et al., 2008; Abe et al., 2024). Such an analysis does not reveal periodic variations of the solar neutrino fluxes but shows how the pdf scaling exponent departs in the non-Gaussian case from the Hurst exponent. Figures 3 and 4 show the scaling exponents (DEA) for the SuperKamiokande I and II data. The respective Hurst exponents for SDA are visible in Figures 5 and 6 (Haubold and Mathai, 2018, 2025). SuperKamiokande is sensitive mostly to neutrinos from the $^8B$ and $hep$ branch of the $pp$ nuclear fusion chain in solar burning. Above approximately 4 $MeV$ the detector can pick-out the scattering of solar neutrinos off atomic electrons which produces Cherenkov radiation in the detector. The $^8B$ and rarer $hep$ neutrinos have a spectrum which ends near 20 $MeV$.

Assuming that the solar neutrino signal is governed by a probability density function with scaling given by the asymptotic time evolution of a pdf of $x$, obeying the property (Scafetta, 2010; Culbreth et al, 2023; Beghin et al. 2025)

$$p(x,t)=\frac{1}{t^\delta}f(\frac{x}{t^\delta}),\eqno(1)$$
where $\delta$ denotes the scaling exponent of the pdf. In the variance based methods, scaling is studied by direct evaluation of the time behavior of the variance of the diffusion process. If the variance scales, one would have

$$\sigma_x^2(t)\sim t^{2H},\eqno(2)$$
where $\sigma_x^2(t)$ is the variance of the diffusion process and where $H$ is the Hurst exponent. To evaluate the Shannon entropy of the diffusion process at time $t$, defined $S(t)$ as

$$S(t)=-\int^{+\infty}_{-\infty} dx\;p(x,t) \ln\;p(x,t)\eqno(3)$$
and with the previous $p(x,t)$ one has

$$S(t)=A+\delta \ln(t),\;\;A=-\int^{+\infty}_{-\infty}dyf(y)\ln f(y).\eqno(4)$$

The scaling exponent $\delta$ is the slope of the entropy against the logarithmic time scale. The slope is visible in Figures 3 and 4 for the SuperKamiokande data measured for $^8B$ and $hep$. The Hurst exponents (SDA) are $H=0.66$ and $H=0.36$ for $^8B$ and $hep$, respectively, see Figures 5 and 6 (Mathai and Haubold, 2018). The $pdf$ scaling exponents (DEA) are $\delta=0.88$ and $\delta=0.80$ for $^8B$ and $hep$, respectively, as shown in Figures 3 and 4. The values for both SDA and DEA indicate a deviation from Gaussian behavior which would require that $H=\delta=0.5$.\\
A test computation for the application of SDA and DEA to data that are known to exhibit non-Gaussian behavior have been published by Haubold et al. (2012) and Tsallis (2024). In this test computation, SDA and DEA, applied to the magnetic field strength fluctuations recorded by the Voyager-I spacecraft in the heliosphere clearly revealed the scaling behavior of such fluctuations as previously already discovered by non-extensive statistical mechanics considerations that lead to the determination of the non-extensivity q-triplet.

\begin{figure}
\resizebox{12cm}{!}{\includegraphics{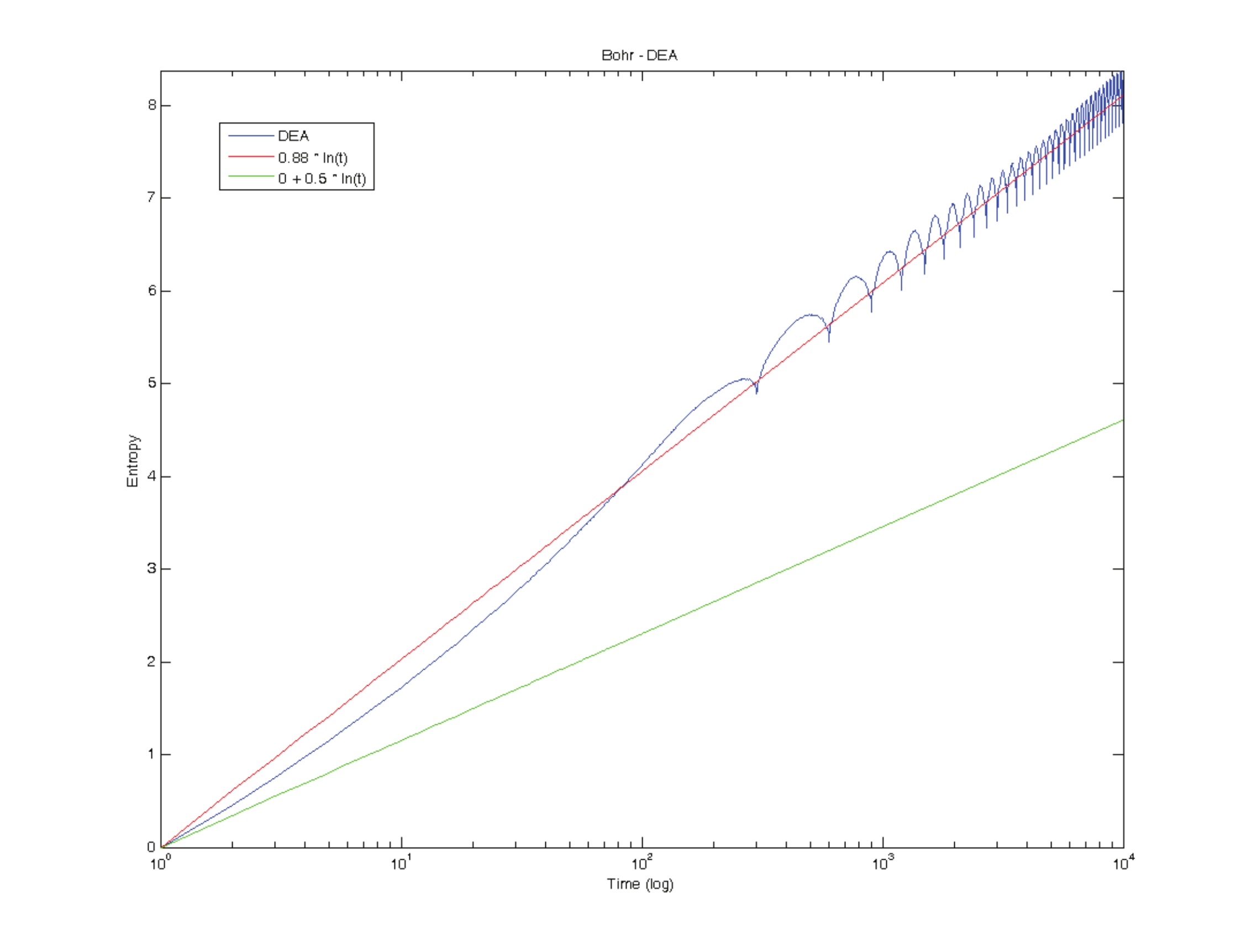}}
\caption{The Diffusion Entropy Analysis (DEA) of the $^8B$ solar neutrino data from the SuperKamiokande I and II experiment.}
\end{figure}

\begin{figure}
\resizebox{12cm}{!}{\includegraphics{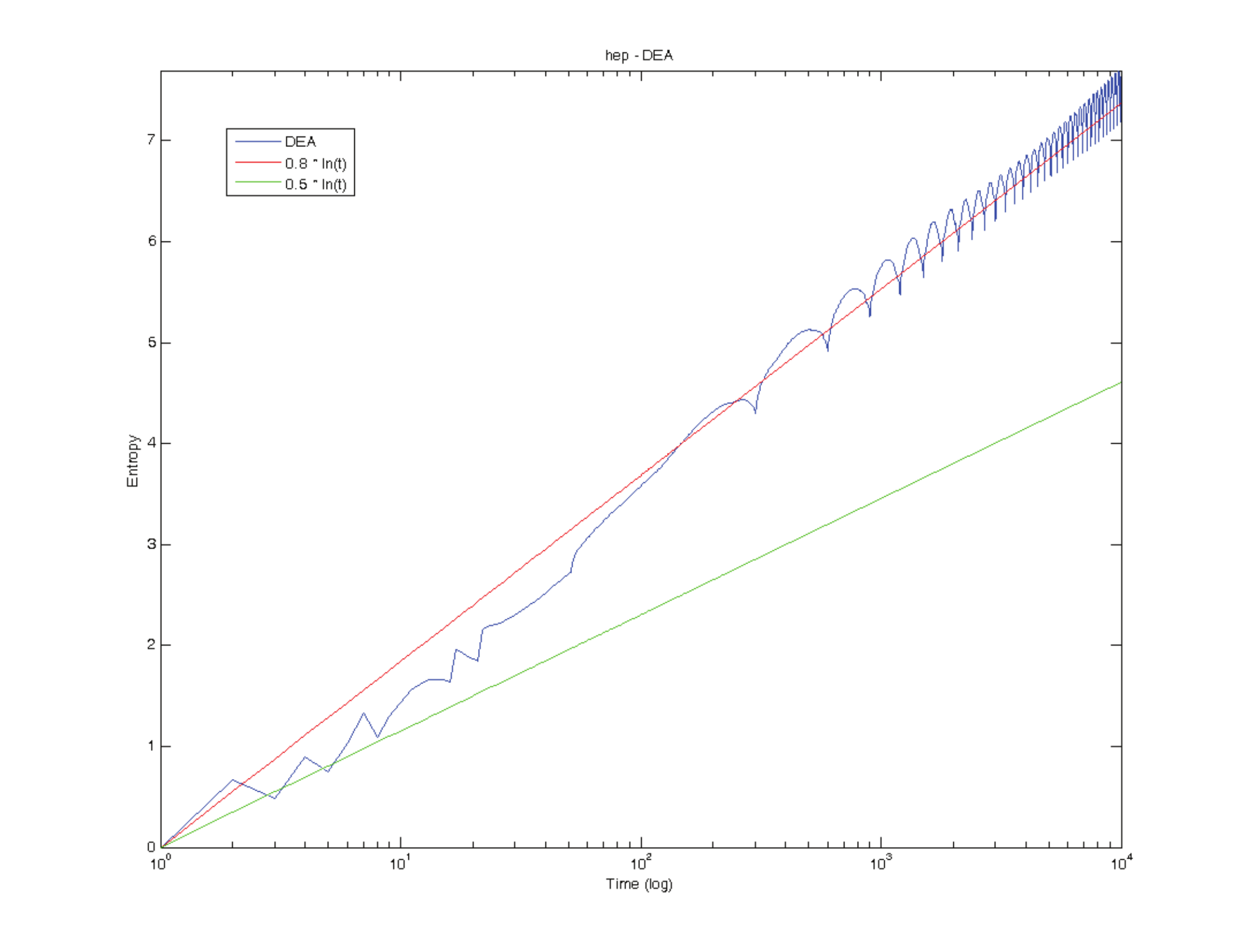}}
\caption{The Diffusion Entropy Analysis (DEA) of the $hep$ solar neutrino data from the SuperKamiokande I and II experiment.}
\end{figure}

\begin{figure}
\resizebox{12cm}{!}{\includegraphics{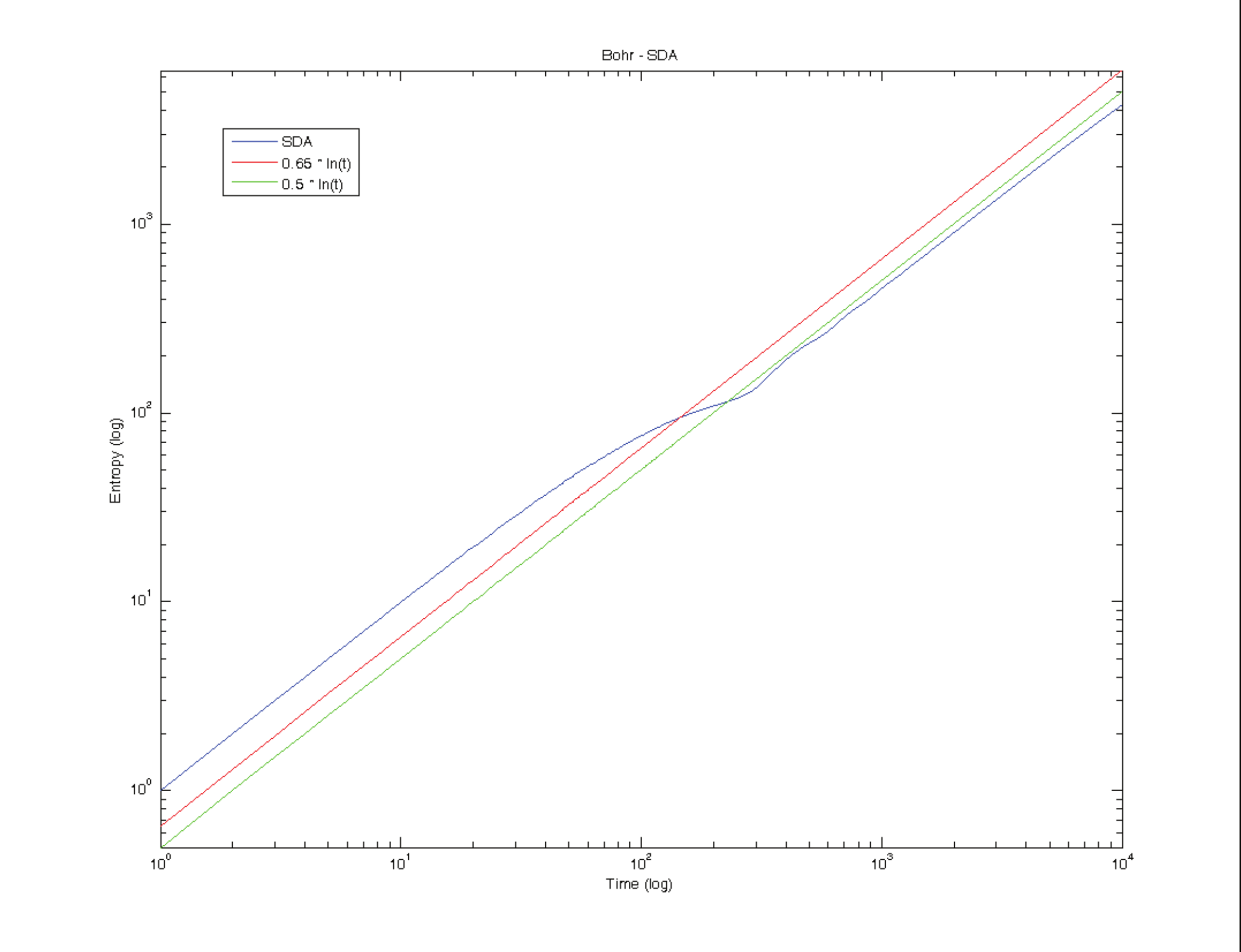}}
\caption{The Standard Deviation Analysis (SDA) of the $^8B$ solar neutrino data from the SuperKamiokande I and II experiment.}
\end{figure}

\begin{figure}
\resizebox{12cm}{!}{\includegraphics{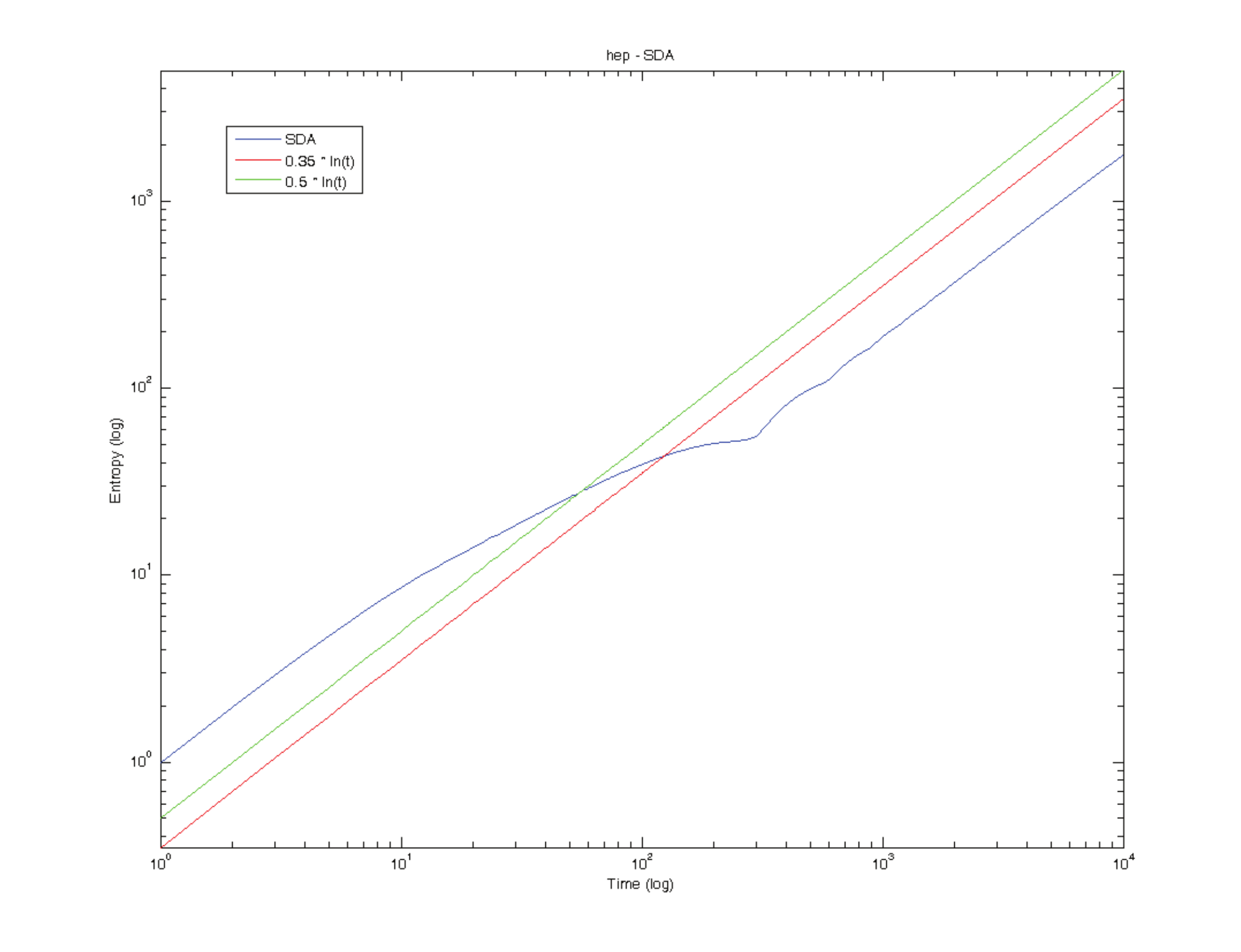}}
\caption{The Standard Deviation Analysis (SDA) of the $hep$ solar neutrino data from the SuperKamiokande I and II experiment.}
\end{figure}

\vskip.3cm\noindent{\bf 2.1\hskip.3cm The Principle of Scaling Model}
\vskip.3cm\noindent
 
Let $x$ be a real scalar random variable having the density $f(x)$. Then $f(x)>0$ on the support of $x$,  that is, whenever $f(x)\ne 0$. Let $a>0$ be a real scalar constant. Suppose that we wish to determine the density of the scaled $x$, namely $ax$. Let $y=ax,a>0$ where $a$ is a positive constant. Let $g(y)$ be the density of $y$. Then 
\begin{align*}
g(y){\rm d}y&=f(x){\rm d}x=f(\frac{y}{a})\frac{{\rm d}y}{a}=\frac{1}{a}f(\frac{y}{a}){\rm d}y\Rightarrow \\
g(y)&=\frac{1}{a}f(\frac{y}{a}).\tag{i}\end{align*}
What is the mean value and variance of the scaled $x$? Let $E[\cdot]$ denote the expected value or mean value of $[\cdot]$ and let ${\rm Var}(\cdot)=$ variance of $(\cdot)$. Then,
$$E[y]=E[ax]=aE[x], {\rm Var}(y)={\rm Var}(ax)=a^2{\rm Var}(x)$$
which means that the mean value is scaled $a$ and the variance is scaled by $a^2$. Suppose that $a=t^{\delta},t>0,\delta>0$. Then, the mean value of $x$ is scaled by $t^{\delta}$ and the variance is scaled by $t^{2\delta}$. What happens to Shannon entropy? Shannon entropy for a discrete distribution $P=(p_1,...,p_k),p_j>0,j=1,...,k,p_1+...+p_k=1$, denoted by $S(P)$, is given by
$$S(P)=-c\sum_{j=1}^kp_j\ln p_j\eqno(ii)$$
where $c>0$ is a constant. This negative sign is taken because $0<p_j<1\Rightarrow \ln p_j<0$ and hence, in order to make $S(P)>0$ a minus sign is taken. The corresponding continuous case is the following:
$$S(f)=-c\int_xf(x)\ln f(x){\rm d}x, c>0.\eqno(iii)$$
Even though $f(x)$ is the density of $x$, it need not remain less than one, still a minus sign is taken because the probability over the interval ${\rm d}x$ is $f(x){\rm d}x$ which is always less than one, there by $\ln [f(x){\rm d}x] <0$ and such sums are there in the integral. Hence the notation $S(f)$ corresponds to that in the discrete case $S(P)$. What happened to Shannon entropy in the scaled case? Note that in the scaled case the density $g(y)$ of the scaled $x$, namely $y=ax$, is given by $g(y)=\frac{1}{a}f(\frac{y}{a}).$ Hence
\begin{align*}
S(g)&=-c\int_yg(y)\ln g(y){\rm d}y=-c[\int_y\frac{1}{a}f(\frac{y}{a})\ln (\frac{1}{a}f(\frac{y}{a})){\rm d}y]\\
&=-c[\ln (\frac{1}{a})]\int_yf(\frac{y}{a})\frac{{\rm d}y}{a} -c\int_yf(\frac{y}{a})\ln f(\frac{y}{a})\frac{{\rm d}y}{a}\\
&=[c\ln a]\int_zf(z){\rm d}z-c\int_zf(z)\ln f(z){\rm d}z, z=\frac{y}{a}\\
&=B+S(f), B=c\ln a.\end{align*}
This is the change where $S(f)$ and $S(g)$ are Shannon  entropy on $f$ and $g$ respectively. If $ a=t^{\delta},t>0,\delta>0 $, then $ B=c\delta \ln t $.

\vskip.2cm
If the staring variable is a $p\times 1$ vector or a $p\times q$ matrix $X$, then the scaling constant can be a real scalar quantity $a>0$ or a real $p\times p$ positive definite matrix $A$ so that $Y=aX$ or $Y=AX$ will be the scaled model. When matrices are involved,  we can have scaling matrices $A^{\frac{1}{2}}$ and $B^{\frac{1}{2}}$ so that we may take $Y=A^{\frac{1}{2}}XB^{\frac{1}{2}}$ where $B$ is another $q\times q$ real positive definite matrix. In the case of matrices or sequence of matrices we can incorporate scaling in many different ways.
\vskip.2cm
In a practical situation, usually the scaling constant $a>0$ may have its own prior distribution. For example, if $a=t^{\delta}$ where if $t$ is the time parameter then the behavior of $y=ax=t^{\delta}x$ may change for different $t$ values or at different epochs of time, which means $t$ has its own distribution.

\vskip.3cm
\noindent
\vskip.3cm\noindent{\bf 2.2\hskip.3cm Prior distribution for the scaling parameter}
\vskip.3cm\noindent
Suppose that the scalar variable $x$ be positive and has a gamma density of the form
$$f(x)=\frac{b^{\gamma}}{\Gamma(\gamma)}x^{\gamma-1}{\rm e}^{-bx},x>0,b>0,\gamma>0$$
and $f(x)=0$ elsewhere. Consider a scaled $x$, namely $y=ax,a>0$. Then, the density of $y$, $g(y)$, is the following:
\begin{align*}
g(y)&=\frac{1}{a}f(\frac{y}{a})=\frac{1}{a}\frac{b^{\gamma}}{\Gamma(\gamma)}(\frac{y}{a})^{\gamma-1}{\rm e}^{-\frac{by}{a}}\\
&=a^{-\gamma}\frac{b^{\gamma}}{\Gamma(\gamma)}y^{\gamma-1}{\rm e}^{-\frac{b}{a}y}.\end{align*}
Suppose that $-\infty< x <\infty, a>0$ and suppose that $a=t^{\delta},t>0,\delta>0 $, then for $y=t^{\delta}x$ the density of $y$ is the following:
$$g(y)=t^{-\gamma\delta}\frac{b^{\gamma}}{\Gamma(\gamma)}y^{\gamma-1}{\rm e}^{-bta^{-\delta}y}.$$
Suppose that $-\infty< x <\infty$ and $x$ has a Gaussian density
$$f(x)=\frac{1}{\sigma\sqrt{2\pi}}{\rm e}^{-\frac{1}{2\sigma^2}(x-\mu)^2}.$$
Then, $y=ax$ has the density
$$g(y)=\frac{1}{a}\frac{1}{\sigma\sqrt{2\pi}}{\rm e}^{-\frac{1}{2\sigma^2a^2}(y-a\mu)^2}.$$
Suppose that the scaling constant $a$ has a prior density. In this case we may write $g(y)$ as $g(y|a)=$ density of $y$ at a given $a$. If $h(a)$ is a prior density for $a$, then the joint density of $a$ and $y$ is $g(y|a)h(a)=\frac{1}{a}f(\frac{y}{a})h(a)$. Then the unconditional density of $y$ or the density of $y$ for all values of $a$, is given by the following and denoted by $g_y(y)$: 
$$g_y(y)=\int_a\frac{1}{a}f(\frac{y}{a})h(a){\rm d}a.$$
Suppose that $h(a)$ is the density
$$h(a)=\frac{1}{\Gamma(\gamma_1)}a^{\gamma_1-1}{\rm e}^{-a},a>0,\gamma_1>0$$
and zero elsewhere. Suppose that $f(x)$ is Gaussian with $\mu=0$. Then,

\begin{align*}
g_y(y)&=\int_a\frac{1}{a}f(\frac{y}{a})h(a){\rm d}a\\
&=\int_{a=0}^{\infty}\frac{1}{a}\frac{1}{\sigma\sqrt{2\pi}}{\rm e}^{-\frac{y^2}{2\sigma^2 a^2}}\frac{1}{\Gamma(\gamma_1)}a^{\gamma_1-1}{\rm d}^{-a}{\rm d}a\\
&=\frac{1}{\sigma\sqrt{2\pi}}\frac{1}{\Gamma(\gamma_1)}\int_{a=0}^{\infty}a^{(\gamma_1-1)-1}{\rm e}^{-a-\frac{y^2}{2\sigma^2a^2}}{\rm d}a.\tag{iv}\end{align*}
This is Bessel type integral which was evaluated earlier. This integral has the structure of the Mellin convolution of a product, namely
$$\int_{v=0}^{\infty}\frac{1}{v}f_1(\frac{u}{v})f_2(v){\rm d}v$$
where
\begin{align*}
f_1(x_1)&={\rm e}^{-x_1^2}\Rightarrow f_1(\frac{u}{v})={\rm e}^{-\frac{u^2}{v^2}},v=a,u^2=\frac{y^2}{2\sigma^2}\\
f_2(x_2)&=x_2^{\gamma_1-1}{\rm e}^{-x_2},x_j>0,j=1,2.\end{align*} Then the Mellin transforms of $f_1$ and $f_2$, denoted by $M_{f_j}(s),j=1,2$, with the Mellin parameter $s$, are the following:

\begin{align*}
M_{f_2}(s) &= \int_0^{\infty}x_2^{s-1}f_2(x_2){\rm d}x_2=\Gamma(\gamma_1-1+s),\Re(s)>-\Re(\gamma_1)+1. \\
M_{f_1}(s)&=\int_{x_1=0}^{\infty}x_1^{s-1}{\rm e}^{-x_1^2}{\rm d}x_1=\frac{1}{2}\Gamma(\frac{s}{2}),\Re(s)>0,\\
M_{f_1}(s)M_{f_2}(s)&=\frac{1}{2}\Gamma(\frac{s}{2} )\Gamma(\gamma_1-1+s).
\end{align*}

Hence, for $u=+\sqrt{\frac{y^2}{2\sigma^2}}$,
\begin{align*}
g_y(y)&=\frac{1}{\sigma\sqrt{2\pi}}\frac{1}{\Gamma(\gamma_1)}\frac{1}{2}\\
&\times \frac{1}{2\pi i}\int_{c-i\infty}^{c+i\infty}\Gamma(\frac{s}{2} )\Gamma(\gamma_1-1+s)u^{-s}{\rm d}s\\
&=\frac{1}{2\sigma\sqrt{2\pi}\Gamma(\gamma_1)}H^{2,0}_{0,2}\left[ u\vert_{(0,\frac{1}{2}),(\gamma_1-1,1)}\right ]
\end{align*}
This H-function, denoted by $H_1$, is the following after writing $\frac{s}{2}=s_1$:
$$H_1=\frac{1}{2\pi i}\int_{c_1-i\infty}^{c_1+i\infty}\Gamma(s_1)\Gamma(\gamma_1-1+2s_1)(u^2)^{-s_1}{\rm d}s_1.$$
By using the duplication formula for gamma functions, $\Gamma(2z)=\pi^{-\frac{1}{2}}2^{2z-1}\Gamma(z)\Gamma(z+\frac{1}{2})$ we can write
$$\Gamma(\gamma_1-1+2s_1)=\frac{2^{\gamma_1-2}}{\pi^{\frac{1}{2}}}(\frac{1}{4})^{-s_1}\Gamma(s_1+\frac{\gamma_1-1}{2})\Gamma(s_1+\frac{\gamma_1}{2}).$$ Then
$$H_1=\frac{2^{\gamma_1-2}}{\pi^{\frac{1}{2}}}G^{3,0}_{0,3}\left[\frac{u^2}{4}\vert_{0,\frac{\gamma_1-1}{2},\frac{\gamma_1}{2}}\right],\frac{u^2}{4}=\frac{y^2}{8\sigma^2}.$$
When $\gamma_1$ is not an odd positive integer, then the poles of the integrand are simple and one can evaluate the above G-function as a linear function of three ${_0F_2}$ hypergeometric functions by using residue calculus as illustrated below for one such series. Consider the poles of $\Gamma(s_1)$. They are at $s_1=-\nu,\nu=0,1,...$. Let the sum of the residues be denoted by $S_1$. Then
$$S_1=\sum_{\nu=0}^{\infty}\frac{(-1)^{\nu}}{\nu!}(\frac{y^2}{2\sigma^2})^{\nu}\Gamma(-\nu+\frac{\Gamma_1-1}{2})\Gamma(-\nu+\frac{\gamma_1}{2}).$$
But for $b\ne \pm  0,1,2,...$
$$\Gamma(b-\nu)=(-1)^{\nu}\frac{\Gamma(b)}{(-b+1)_{\nu}},(d)_{\nu}=d(d+1)...(d+\nu-1),(d)_0=1,d\ne 0.$$

Then,
\begin{align*}
\Gamma(-\nu+\frac{\gamma_1-1}{2})&=(-1)^{\nu}\Gamma(\frac{\gamma_1-1}{2})\frac{1}
{(\frac{3}{2}-\frac{\Gamma_1}{2})_{\nu}}\\
\Gamma(-\nu+\frac{\gamma_1}{2})&=(-1)^{\nu}\Gamma(\frac{\gamma_1}{2})\frac{1}{(1-\frac{\gamma_1}{2})_{\nu}}.\end{align*}
Hence,
$$S_1=\Gamma(\frac{\gamma_1}{2})\Gamma(\frac{\gamma_1-1}{2}){_0F_2}(-;\frac{3}{2}-\frac{\gamma_1}{2},1-\frac{\gamma_1}{2};-\frac{y^2}{8\sigma^2}). $$

\vskip.3cm\noindent{\bf 3.\hskip.3cm Probability Density Function and Differential Equation: L\'evy Flights}
\vskip.3cm\noindent
We consider a diffusion process generated by a waiting time pdf with a finite characteristic time $T$ that can be modeled with $a$ Poissonian distribution, and a jump length pdf
$\lambda (x)$ given by a L\'evy distribution with index $0<\alpha<2$ (Metzler and Klafter, 2000). The Fourier transform of $\lambda(x)$ is

$$\hat{\lambda}(k)=exp(-\sigma^\alpha|k|^\alpha)\sim 1-\sigma^\alpha|k|^\alpha.\eqno(5)$$
Then
$\lambda(x)$ has the asymptotic behavior given by

$$\lambda(x) \sim A_\alpha \sigma^\alpha|x|^{-1-\alpha}=A_\alpha \sigma^{1-\mu}|x|^{-\mu}\eqno(6)$$
for  $|x|\gg \sigma$ and $\mu=1+\alpha$. Substituting the asymptotic expansion of the jump length pdf
$\hat{\lambda}(k)$ in the Fourier space and the waiting time pdf of

$$\Phi(t)=\frac{1}{\tau} exp(-\frac{t}{\tau})\eqno(7)$$
where $\tau=T<\infty$ is the characteristic waiting time in the Laplace space into

$$\hat{p}(k,s)=\frac{1-\hat{\Phi}(s)}{s}\frac{\hat{p_0}(k)}{1-\hat{\Phi}(s)\hat{\lambda}(k)},\eqno(8)$$
where $\hat{p_0}(k)$ is the Fourier transform of the initial condition $p(x,0)$, we obtain the following jump pdf in the Fourier-Laplace space

$$\hat{p}(k,s)=\frac{1}{s+K^\alpha|k|^\alpha},\eqno(9)$$
where $K^\alpha = \sigma^\alpha/\tau$ is the generalized diffusion constant. Eq.(9) is the solution of the generalized diffusion equation (Hilfer, 2018)

$$\frac{\partial p(x,t)}{\partial t}= K^\alpha\;_{-\infty}D^\alpha_x p(x,t),\eqno(10)$$
where $_{-\infty}D^\alpha_x$ is the fractional Weyl operator.
Upon Laplace inversion of Eq. (9), we get the characteristic function of the jump pdf

$$\hat{p}(k,t)= exp(-K^\alpha t|k|^\alpha).\eqno(11)$$
Eq.(11) is the characteristic function of a centered and symmetric L\'evy distribution. The Fourier inversion of (7) can be analytically obtained by making use of the Fox function (Mathai et al., 2010; Mathai and Haubold, 2018)

\begin{align*}
p(x,t)&=\frac{1}{\alpha}\frac{1}{t^{1/\alpha}}\frac{t^{1/\alpha}}{|x|}H_{2,2}^{1,1}\left[\frac{|x|}{Kt^{1/\alpha}}\big\vert_{(1,1),(1,\frac{1}{2})}^{(1,\frac{1}{\alpha}),(1,\frac{1}{2})}\right]\\
&=\frac{1}{\alpha |x|}\frac{1}{2\pi i}\int_{c-i\infty}^{c+i\infty}\frac{\Gamma(1+\frac{s}{\alpha})\Gamma(-\frac{s}{2})}{\Gamma(-s)\Gamma(1+\frac{s}{2})}\left(\frac{|x|}{Kt^{1/\alpha}}\right)^{-s}{\rm d}s\end{align*},
where $c$ in the contour is such that $-\alpha<c<0$. Replacing $\frac{s}{\alpha}$ by $s$, observing that $\alpha$ coming from $s~{\rm d}s$ is canceled with $\alpha$ sitting outside, we have the following:

$$p(x,t)=\frac{1}{|x|}\frac{1}{2\pi i}\int_{c'-i\infty}^{c'+i\infty}\frac{\Gamma(1+s)\Gamma(-\frac{\alpha}{2}s)}{\Gamma(-\alpha s)\Gamma(1+\frac{\alpha}{2}s)}\left(\frac{|x|}{Kt^{1/\alpha}}\right)^{-\alpha s}{\rm d}s,-1<c'<0.$$
By using the duplication formula for gamma functions, we have
$$\Gamma(-\alpha s)=\Gamma(2(-\frac{\alpha}{2}s))=2^{-\alpha s-1}\pi^{-\frac{1}{2}}\Gamma(-\frac{\alpha}{2}s)\Gamma(\frac{1}{2}-\frac{\alpha}{2}s)$$
so that one $\Gamma(-\frac{\alpha}{2}s)$ is canceled. Then,
$$p(x,t)=\frac{2\pi^{\frac{1}{2}}}{|x|}\frac{1}{2\pi i}\int_{c'-i\infty}^{c'+i\infty}\frac{\Gamma(1+s)}{\Gamma(\frac{1}{2}-\frac{\alpha}{2}s)\Gamma(1+\frac{\alpha}{2}s)}\left(\frac{|x|}{2Kt^{1/\alpha}}\right)^{-\alpha s}{\rm d}s.$$
Evaluating the H-function as the sum of the residues at the poles of $\Gamma(1+s)$, which are at $s=-1-\nu,\nu=0,1,....$, we have the following series:
\begin{align*}
p(x,t)&=\frac{2\pi^{\frac{1}{2}}}{|x|}(\frac{|x|^{\alpha}}{(2K)^{\alpha}t})\\
&\times \sum_{\nu=0}^{\infty}\frac{(-1)^{\nu}}{\nu!}\frac{1}{\Gamma(\frac{1+\alpha}{2}+\frac{\alpha}{2}\nu)\Gamma(1-\frac{\alpha}{2}(1+\nu))}\left(\frac{|x|^{\alpha}}{(2K)^{\alpha}t}\right)^{\nu},\alpha\ne 1,\alpha\ne 2.
\end{align*}

\vskip.3cm\noindent{\bf 4.\hskip.3cm Discussion}
\vskip.3cm\noindent
The first solar neutrino experiment led by Raymond Davis Jr. showed a deficit of neutrinos relative to the solar model prediction, referred to as the solar neutrino problem since the 1970s. The Kamiokande experiment led by Masatoshi Koshiba successfully observed solar neutrinos, as first reported in 1980s. The solar neutrino problem was solved due to neutrino oscillations by comparing the SuperKamiokande and Sudbury Neutrino Observatory results. While recent decades have offered tremendous advances in solar neutrinos across the fields of solar physics (Buldgen et al., 2024; Yang and Tian, 2024), nuclear physics (Bertulani et al., 2022; Hwang et al., 2023), and neutrino physics (Sturrock et al., 2021; Slad, 2024), many lingering mysteries remain.\\
This chapter takes advantage of publicly available solar neutrino SuperKamiokande data and analyses them by applying diffusion entropy analysis and standard deviation analysis. The result is a scaling exponent $\delta<1$ indicating anomalous diffusion of solar neutrinos in terms of L\'evy flights. Based on this result the chapter developes the probability density function for neutrino flights and derives the respective differential equation in terms of a fractional Fokker-Planck equation. Accordingly, the closed form analytic representation of the neutrino power density function is given as a Fox H-function that can be used for further numerical exercises for the benefit of solar neutrino physics.\\

The authors have grateful for the support in the diffusion entropy analysis and standard deviation analysis by Dr. Alexander Haubold while he was doing his research at the Department of Computer Science, Columbia University, New York (USA). The results of diffusion entropy analysis were independently confirmed by the research of Dr. Nicy Sebastian, Department of Statistics, St Thomas College, Thrissur, University of Calicut, Kerala (India).\\

\vskip.3cm\noindent\begin{center}{\bf References}\end{center}

\vskip.3cm\noindent Abe, K. et al. 2023. Search for periodic time variations of the solar $^8B$ neutrino flux between 1996 and 2018 in SuperKamiokande.
 https://arxiv.org/pdf/2311.01159.

\vskip.3cm\noindent Abe, K. et al. 2024. Solar neutrino measurements using the full data period of SuperKamiokande-IV. https://doi.org/10.1103/PhysRevD.109.092001.

\vskip.3cm\noindent Beghin, L., Cristofaro, L., and da Silva, J.L. 2025. Fox-H densities and completely monotone generalized Wright functions. Journal of Theoretical Probability 38: 18. https://doi.org/10.1007/s10959-024-01391-9.

\vskip.3cm\noindent Bertulani, C.A., Hall, F.W., and Santoyo, B.I. 2022. Big Bang nucleosynthesis as a probe of new physics. https://arxiv.org/abs/2210.04071.

\vskip.3cm\noindent Buldgen, G., Noels, A., Scuflaire, R., Amarsi, A.M., Grevesse, N., Eggenberger,P.,Colgan, J., Fontes, C.J., Baturin, V.A., Oreshina, A.V., Ayukov, S.V., Hakel, P., and Kilcrease, D.P.  In-depth analysis of solar models with high-metallicity abundances and updated opacity tables. https://arxiv.org/abs/2404.10478.

\vskip.3cm\noindent Cravens, J.P. et al. 2008. Solar neutrino measurements in SuperKamiokande-II. Physical Review D 78: 032002.

\vskip.3cm\noindent Culbreth, G., Baxley, J., and Lambert, D. 2023. Detecting temporal scaling with modified diffusion entropy analysis. https://arxiv.org/abs/2311.11453.

\vskip.3cm\noindent Haubold, A., Haubold, H.J., and Kumar, D. 2012. Heliosheath: Diffusion entropy analysis and nonextensivity q-triplet. https://arxiv.org/abs/1202.3417v1.

\vskip.3cm\noindent Hilfer, R. 2018. Mathematical and physical interpretation of fractional derivatives and integrals. In Handbook of Fractional Calculus with Applications, Volume 1: Basic Theory, Edited by A. Kochubei and Yu. Luchko.  Berlin: de Gruyter, pp. 47-85.

\vskip.3cm\noindent Hwang, E., Ko, H., Heo, K., Cheoun, M.-K., and Jang, D. 2023. Revisiting the Gamow factor reactions on light nuclei. https://arxiv.org/abs/2302.10102.

\vskip.3cm\noindent Mathai, A.M., and Haubold, H.J. 2018. Erd\'elyi-Kober Fractional Calculus: From a Statistical Perspective, Inspired by Solar Neutrino Physics. Singapore: Springer.

\vskip.3cm\noindent Mathai, A.M., Saxena, R.K., and Haubold, H.J. 2010. The H-Function: Theory and Applications. New York: Springer.

\vskip.3cm\noindent Mathai, A.M., and Haubold, H.J. 2025. Modified models for neutrino masses and mixings. Journal of Mathematical Physics 66, 083506. doi: 10.1063/5.0266171.

\vskip.3cm\noindent Metzler, R., and Klafter, J. 2000. The Random Walk's Guide to Anomalous Diffusion: A Fractional Dynamics Approach. Physics Reports 339: 1-77.

\vskip.3cm\noindent Orebi Gann, G.D., Zuber, K., Bemmerer, D., and Serenelli, A. 2021. The Future of Solar Neutrinos. Annual Review of Nuclear and Particle Science 71: 491-528.

\vskip.3cm\noindent Sakurai, K. 2018. Solar Neutrino Problems - How They Were Solved. Tokyo: TERRAPUB.

\vskip.3cm\noindent Scafetta, N. 2010. Fractal and Diffusion Entropy Analysis of Time Series: Theory, concepts, applications and computer codes for studying fractal noises and L\'{e}vy walk signals. Saarbruecken: VDM Verlag Dr. Mueller.

\vskip.3cm\noindent Slad, L.M. 2024. Logic and numbers related to solar neutrinos. Annalen der Physik (Berlin) 536: 2400168.

\vskip.3cm\noindent Sturrock, P.A., Piatibratova, O., and Scholkmann, F. 2021. Comparative analysis of SuperKamiokande solar neutrino measurements and geological survey of Israel radon decay measurements. Frontiers in Physics 9: 718306.\\
https://doi.org/10.3389/fphy.2021.718306.

\vskip.3cm\noindent Tsallis, C. 2024. Reminiscenses of half a century of life in the world of theoretical physics. Entropy 26: 158. http://doi.org/10.3390/e26020158.

\vskip.3cm\noindent Yang, W., and Tian, Z. 2024. Solar models and astrophysical S-factors constrained by helioseismic results and updated neutrino fluxes. https://arxiv.org/abs/2405.10472.

\vskip.3cm\noindent Yoo J. et al. 2003. Search for periodic modulations of the solar neutrino flux in SuperKamiokande-I. Physical Review D 68: 092002.

\vskip.5cm\noindent{\bf Declarations:}
Ethics, Consent to Participate, and Consent to Publish declarations: not applicable.

\end{document}